\documentclass[aps,prl,preprint,groupedaddress,superscriptaddress]{revtex4}
\usepackage{epsfig,subfigure,subfloat,graphicx,amssymb,latexsym,amsthm
,amsmath,natbib,slashed,dsfont,color,nicefrac}

\newcommand{\beq}{\begin{equation}}
\newcommand{\eeq}{\end{equation}}
\newcommand{\beqa}{\begin{eqnarray}}
\newcommand{\eeqa}{\end{eqnarray}}

\newcommand{\ben}{\begin{enumerate}}
\newcommand{\een}{\end{enumerate}}
\newcommand{\bit}{\begin{itemize}}
\newcommand{\eit}{\end{itemize}}
\newcommand{\bpm}{\begin{pmatrix}}
\newcommand{\epm}{\end{pmatrix}}

\newcommand{\pdn}[3]{\frac{ {\partial}^{#1} #2}{{\partial} #3 ^{#1}}}

\newcommand{\ra}{\rangle}
\newcommand{\la}{\langle}

\def\A{{\cal A}}

\def\H{{\cal H}}

\def\lt{\left}
\def\rt{\right}

\def\bracket#1#2 {\mathinner{\langle{#1}|{#2}\rangle}}

\def \im {{\cal I}m}
\def \re {{\cal R}e}

\def \vphi {\varphi}

\def \Dl {\Delta}
\def \dl {\delta}

\def \gm {\gamma}

\def \sg {\sigma}

\def \th {\theta}
\def \nb {\nabla}

\def\->{\rightarrow}
\def \sr{Schr\"{o}dinger~}

\def\bk{{\bs k}}
\def\bn{{\bs n}}

\def\bp{{\bs p}}

\def \Hm {Hamiltonian}
\def \bs {\boldsymbol}

\def \. {\cdot}
\def \nf{\nicefrac}

\def \. {\cdot}

\begin{document}


\title{Prefect Klein tunneling in anisotropic graphene-like photonic lattices}


\author{Omri Bahat-Treidel}
\affiliation{Department of Physics, Technion-Israel Institute of
Technology, Technion City, Haifa 32000, Israel}
\author{Or Peleg}
\affiliation{Department of Physics, Technion-Israel Institute of
Technology, Technion City, Haifa 32000, Israel}
\author{Mark Grobman}
\affiliation{Department of Physics, Technion-Israel Institute of
Technology, Technion City, Haifa 32000, Israel}
\author{Nadav Shapira}
\affiliation{Department of Physics, Technion-Israel Institute of
Technology, Technion City, Haifa 32000, Israel}
\author{T. Pereg-Barnea}
\affiliation{Department of Physics,
California Institute of Technology, 1200 E. California Blvd, MC114-36,
Pasadena, CA 91125}
\author{Mordechai Segev}
\affiliation{Department of Physics, Technion-Israel Institute of
Technology, Technion City, Haifa 32000, Israel}


\today

\begin{abstract}  
We study the scattering of waves off a potential step in deformed honeycomb lattices.
For small deformations below a critical value, perfect Klein tunneling is obtained. This means that a potential step in any direction transmits waves at normal incidence with unit transmission probability, irrespective of the details of the potential. Beyond the critical deformation, a gap in the spectrum is formed, and a potential step in the deformation direction reflects all normal-incidence waves, exhibiting a dramatic transition form unit transmission to total reflection.
These phenomena are generic to honeycomb lattice systems, and apply to electromagnetic waves in photonic lattices, quasi-particles in graphene, cold atoms in optical lattices.
\end{abstract}


\maketitle



Scattering of relativistic fermions is fundamentally different from
that of non-relativistic ones, since relativistic fermions are
described by the Dirac equation which is first order in momentum (rather than the
second order \sr equation). Relativistic fermions (massive or massless)
incident normally upon a potential step of height $V_0$, exhibit
non-zero transmission probability, even when their energy is smaller
than $V_0$ \cite{klein}. This behavior of relativistic fermions is called Klein tunneling and stands in sharp contrast to the more intuitive result of quantum mechanics for non-relativistic particles,
where the transmission probability vanishes completely when the height of the step is greater than the particle's energy. Outside the step, the state is of positive energy, whereas inside the step there is a propagating negative energy state. This unique scattering process, has never been experimentally
verified, since an experiment designed to observe Klein tunneling with elementary particles requires high fields which are not currently available.
However, it has been predicted that the charge carriers in graphene, that obey the massless Dirac's equation,
exhibit similar behavior \cite{Ando}. More specifically, it has been suggested that
charge carriers in graphene experience non-resonant unit
transmission in monolayer graphene, and total reflection in bilayer
graphene \cite{Katsnelson_b}. Experiments with bipolar junctions were able to show very high
conductance in the presence of a gate voltage, indicating high
transmission probability \cite{Huard-2007-98,stander:026807,kim}.
In addition, unusual transmission properties are predicted in honeycomb photonic crystals \cite{sepkhanov:063813,sepkhanov:045122}.

Here we study the dynamics of waves in a deformed honeycomb photonic lattice, and in particular the tunneling process into a refractive index step. We find that, up to the critical deformation in which a gap in the spectrum is formed, {\it non-resonant unit transmission} is obtained at normal incidence, irrespective of the details of the potential. That is, surprisingly, deformed honeycomb lattices also display perfect tunneling as well as non-deformed honeycomb lattices. At deformations stronger than the critical one, we find {\it non-resonant total reflection} in the deformation direction. Generally, optical structures exhibiting unit-transmission are resonant,
and are characterized by transmission peaks, e.g., Fabry-Perot etalon.
Thus, our study introduces a new domain of light transport in photonic structures, displaying
non-resonant effects, and offering an opportunity to directly observe Klein tunneling.

\begin{figure}[b]
    \center
    {\includegraphics[width=0.50\textwidth]{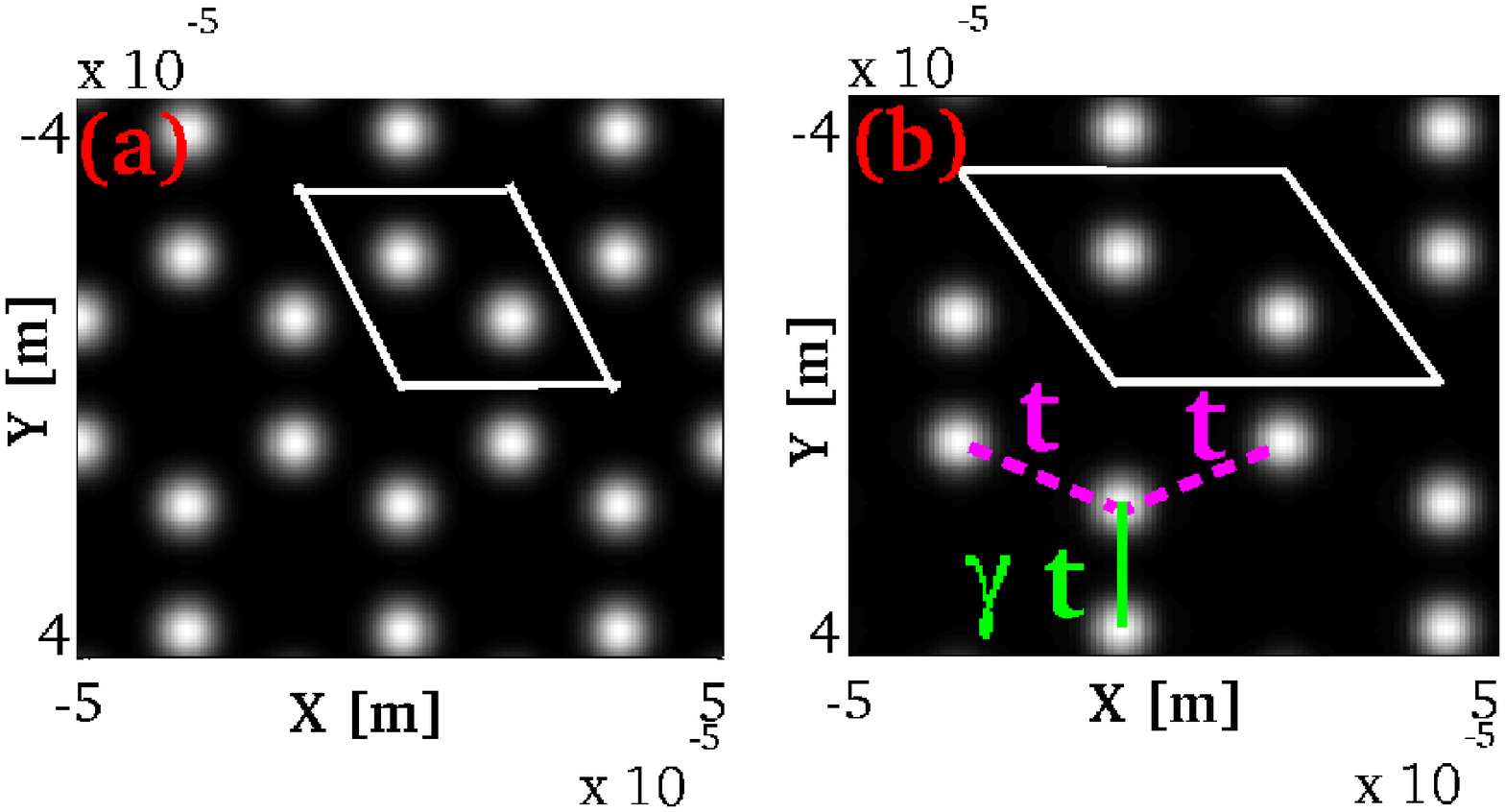}
    \includegraphics[width=0.27\textwidth]{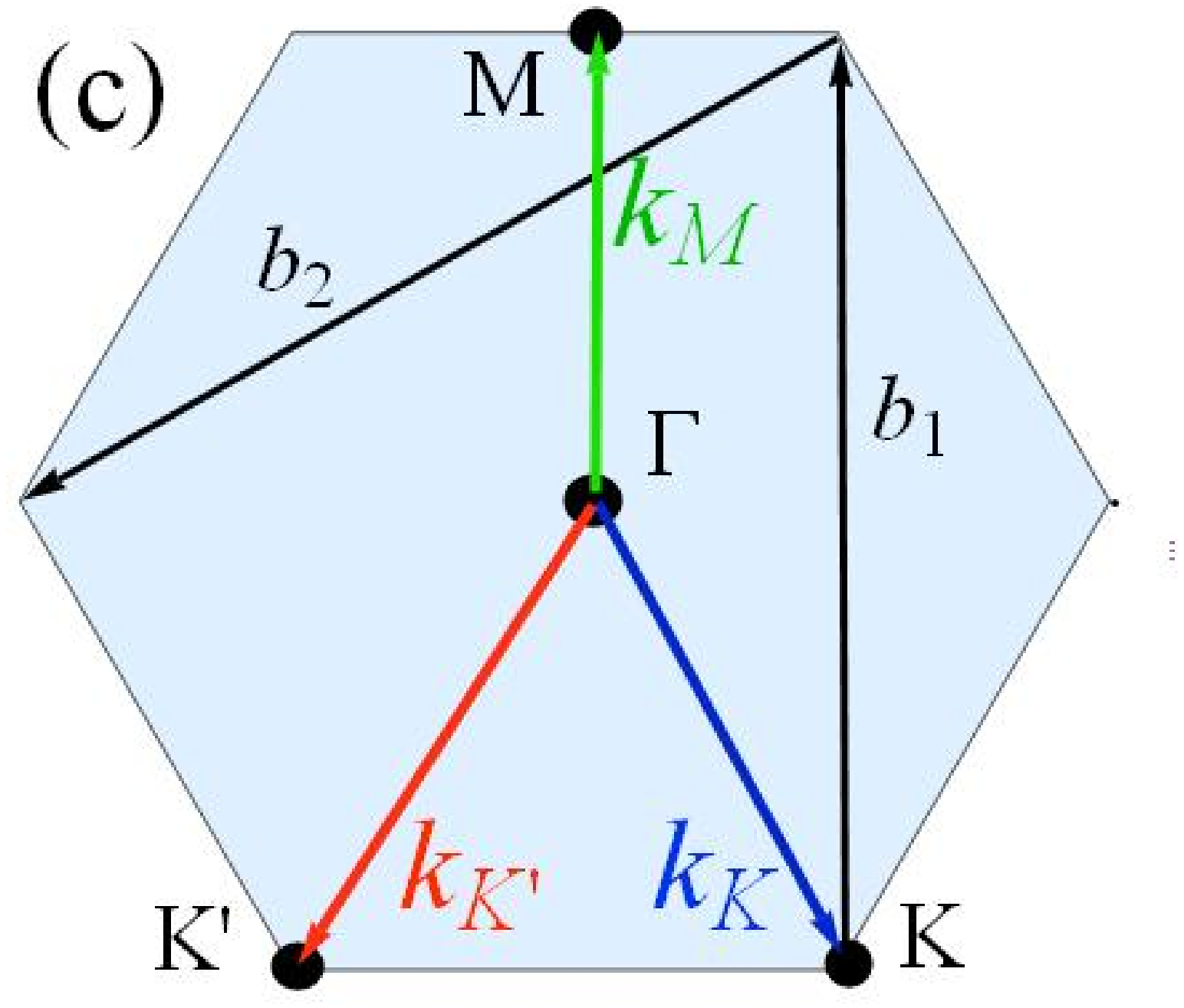}
    }%
    \caption {Non deformed (a) and deformed (b) honeycomb lattices, that have two sites in a unit cell (white).
    (c) The first Brillouin zone with the high symmetry points.
	}
    \label{lattice}
\end{figure}%

Paraxial propagation of a monochromatic field envelope
$\psi$ with a honeycomb refractive index is given by \cite{peleg:103901}%
\beq
    i \pdn{}{\psi}{z} = -\frac{1}{2k}\nb^2_{\perp}\psi -
    \frac{k\dl n(x,y)}{n_0} \psi \equiv \hat H \psi,
    \label{nls}
\eeq %
where $\dl n$ is the modulation in the refractive index, $k$ is the wave-number and $n_0$ the
background refractive index. Note that ${k}\dl n(x,y)/n_0$ is an optical potential with an opposite sign, which means that light is attracted to higher index of refraction. Since $\dl n$ is independent
of $z$, Eq.(\ref{nls}) has solutions of the form $\psi(x,y,z) = U(x,y) \exp(i \beta
z)$, where $\beta$ is the propagation constant, and $U$ is a
solution of $\hat H U = \beta U $, hence $\beta$ is analogous to the energy.
Like other periodic systems, the system can be described by tight binding (TB) and hence the analogy with graphene.
The honeycomb lattice is comprised of two triangular sub-lattices, denoted by A
and B (Fig.\ref{lattice}a). Therefore, when writing the \Hm~in Wannier basis, one must consider
two sets of amplitudes, $a_{\bs n}, b_{\bs n}$, associated with
the two sites $A$ and $B$ in each unit cell located at $\bn$.
Assuming only nearest neighbors hopping (coupling), the TB \Hm~resulting from Eq.(\ref{nls})
reads \cite{ablowitz:053830}
\beq%
    H_0 = -\sum_{\bs n, j} t_j
    \lt(a^{*}_{\bs n} b_{\bs n+\bs \dl_j}
    + b_{\bs n + \bs \dl_j}^{*} a_{\bs n} \rt), \label{H0_a}
\eeq%
where $t_j$'s are the hopping parameters, and $\bs \dl_j$ are the vectors connecting the nearest neighbors.
The anisotropy is manifested by different
hopping parameters in different directions, and can be realized by various means \cite{zhu-2007-98,Bahat-Treidel:08,pereira:045401}. We consider a photonic lattice with different spacing between the sites (Fig.\ref{lattice}b). Such deformations are uniaxial, i.e., $t_2=t_3 = t$ and $t_1 = \gm t$,
where $\gm > 1$. We refer to the $x-$direction as the deformation direction. Expending $a_{\bn},~b_{\bn}$ in Fourier space and defining $\vphi_{\bk} \equiv \sum_j \exp(i \bs \dl_j \bk)$, we can write
\beq
    H_0 =  \frac{1}{N}\sum_{BZ} \Psi^{\dag} \H_k \Psi, \quad
    \H_k \equiv \sg_x \re\{ \vphi_{\bk}\} + \sg_y \im \{\vphi_{\bk}\},
         \label{H0_b}
\eeq%
where $\sigma_i$ are Pauli matrices and $\Psi^{\dag} = ( a_{\bk}^* ~ b_{\bk}^* )$ is a pseudospinor.
The spectrum is obtained by the eigenvalues of $\H_k$
yielding positive and negative branches:
\beqa
    \beta =
    \pm t \sqrt{2 + \gm ^2 + 4 \gm \cos \tfrac{k_x a}{2}
    \cos \tfrac{\sqrt{3} k_y a }{2}  +2 \cos k_x a
    }, \label{deformed analitic band}
\eeqa%
where $a$ is the lattice constant.
For $\gm < 2$, the two branches intersect at two inequivalent points in the first Brillouin zone, known as the Dirac points. The vicinity of these points are the 'valleys', which serve as an additional degree of freedom for excitations with momentum close to the Dirac point (low energy excitations): two are associated with each band and two are associated with each valley.
At $\gm = 2$, the Dirac points merge at the $M-$point (Fig.\ref{lattice}c), and for $\gm > 2$ a gap forms \cite{zhu-2007-98,Wunsch,Bahat-Treidel:08,pereira:045401}. An effective \Hm~is obtained by expanding
$\vphi_{\bk}$ around the extrema of the bands.  Its form depends strongly on the strength of the deformation. For $ 1 \leq \gm < 2$, the effective \Hm~is an anisotropic
Dirac's \Hm~\cite{zhu-2007-98}
\beq
	\H_1 = v_x p_x \sg_x + v_y p_y \sg_y,
\eeq
where $v_y = \sqrt{3}\gm t a/2, ~v_x = t a\sqrt{1-\gm^2/4}$, and $\bp$ is the momentum measured from
the extrema of the bands. As $\gm$ approaches $2$, $v_x$ vanishes
and one must include high order terms in $p_x$. Therefore, at the critical deformation $(\gm = 2)$, the \Hm~has no linear term in $p_x$ \cite{dietl:236405}
\beq
	\H_2 = -\sqrt{3}t a p_y \sigma_y  + \tfrac{1}{4}t a^2 p_x^2 \sg_x.
    \label{H gm=2}
\eeq
Deriving the effective \Hm~for stronger deformations that are characterized by $\gm > 2$, we find
\beq
	\H_3 = \lt[\Dl + \tfrac{ta^2}{4}p_x^2-
    \tfrac{(8\gm-1)ta^2}{12}p_y^2\rt]\sigma_x
    -\tfrac{(2\gm-1)ta}{\sqrt{3}} \sigma_y p_y,
    \label{H gm>2}
\eeq
where $\Dl = t(\gm-2)$.
Note that the quadratic term in $p_y$ may not be neglected compared to the linear term, since the dispersion obtained from (\ref{H gm>2}) must coincide with the expansion of (\ref{deformed analitic band}). We also note that when the Dirac points merge, the valley degree of freedom vanishes and the number of degrees of freedom is reduced to two.

In all three cases, the effective \Hm~has the general form $g(\bs p)\sigma_x + h(\bs p)\sigma_y$,
where $g,h$ are functions of $\bs p$. Defining $F(\bp) \equiv g(\bp) - i h(\bp)$, the eigenstates are%
\beq
	\chi^{(\pm)}(\bs p) = \nf{1}{\sqrt{2}}\bpm 1 & ~\pm F/|\beta| \epm,~~~ \textrm{where}~~~ |F/\beta| = 1,
\eeq%
and '$\pm$' indicate the sign of the propagation constant. In real space,~
$
	\psi^{(\pm)}_{\bp}(x,y) = \chi^{(\pm)}(\bs p) e^{\pm i(p_x x + p_y y)}.
$
\subparagraph{Scattering:}
In order to study the scattering problem, we consider a honeycomb lattice with additional refractive index step with a corresponding optical potential, $V_N$. Eq.(\ref{nls}) then transforms: $H \-> H - V_N$.
We study the scattering of a wave packet from the second band with $\beta = -\beta_0$, that is initially located at the region of higher index, and is traveling towards the interface. The height of the step, $V_0$, is greater than
$\beta_0$, mimicking the scenario considered by Klein. We consider cases: $(i)$ the step is along the direction of the {\it larger} hopping parameter ($y-$direction), and $(ii)$, the step is in the perpendicular direction ($x-$direction). In both scenarios, we calculate the transmission probability below and above the critical deformation, and find it to be qualitatively different. We emphasize that below the critical deformation,
the system is described by $\H_1$, therefore there is no qualitative difference between different directions, i.e., the direction of the step is insignificant.

\subparagraph{Step in $y$:}The additional optical potential is $V_N (x,y) = V_0 \. \Theta(y)$, where $\Theta$ is the Heaviside function. The transmission (reflection) probability, $T$ ($R$), is given by the ratio of the transmitted (reflected) current and the incident current:
\beq
	R = |\A_r |^2, \quad  T = \frac{q_y \beta_0}{p_y(V_0 - \beta_0)}|{\A_t} |^2,
\eeq
where $\A_t~(\A_r)$ is the transmission (reflection) amplitude obtained from continuity of $\psi$
at the boundary:
\beq
	\chi_{-}(p_x,-p_y) + \A_r \. \chi_{-}(p_x,p_y) = \A_t \. \chi_{+}(p_x,q_y), \label{continuity}
\eeq
where $q_y = [(V_0 - \beta_0)^2 + v_{x}^2 p_x^2]^{1/2}$. The negative momentum in $\chi_-$ is
due to the fact that states from the second band have momentum opposite to their velocity.
\begin{figure}[t]
    \center
    {\includegraphics[width=0.48\textwidth]{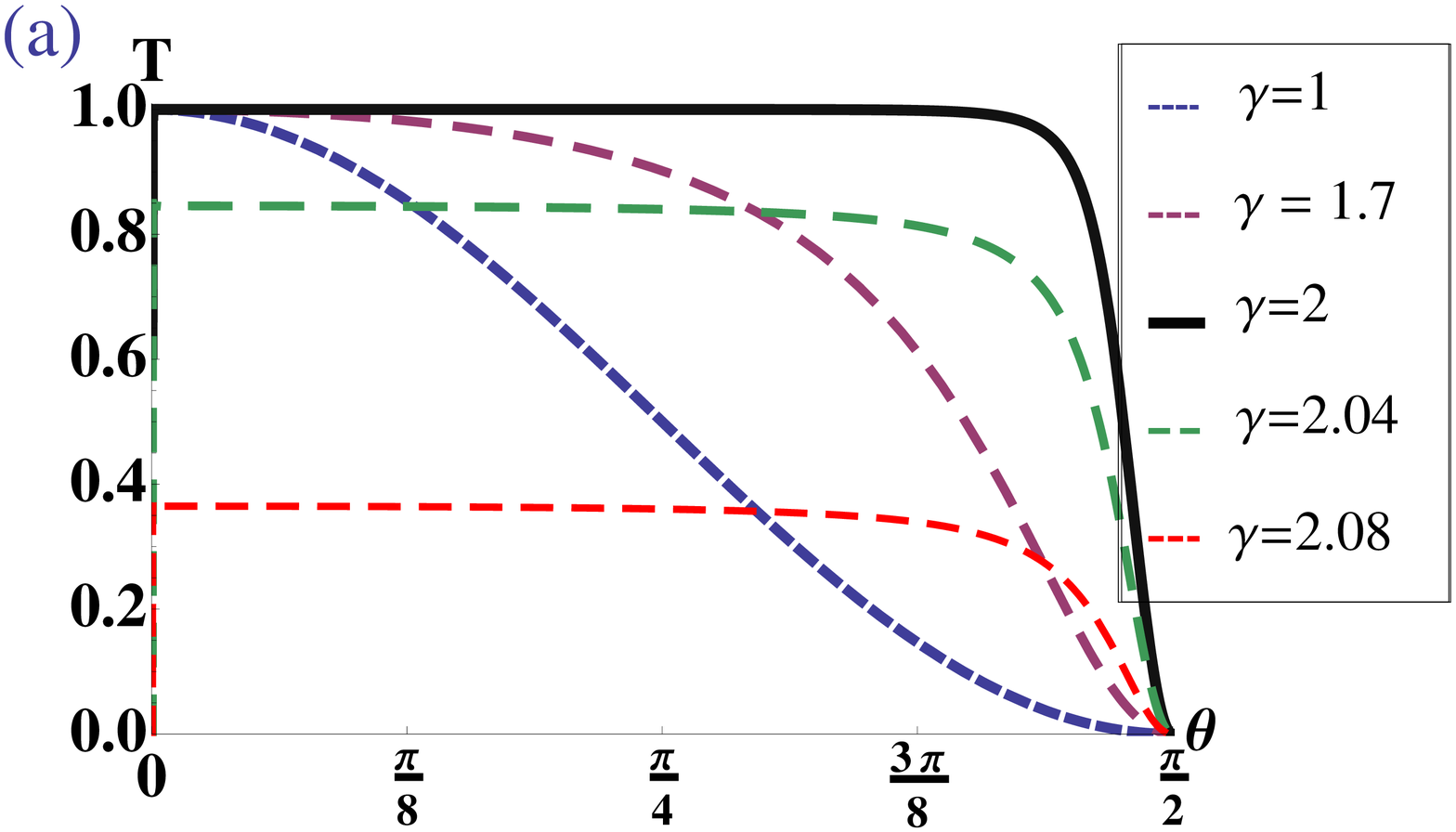}
    \includegraphics[width=0.48\textwidth]{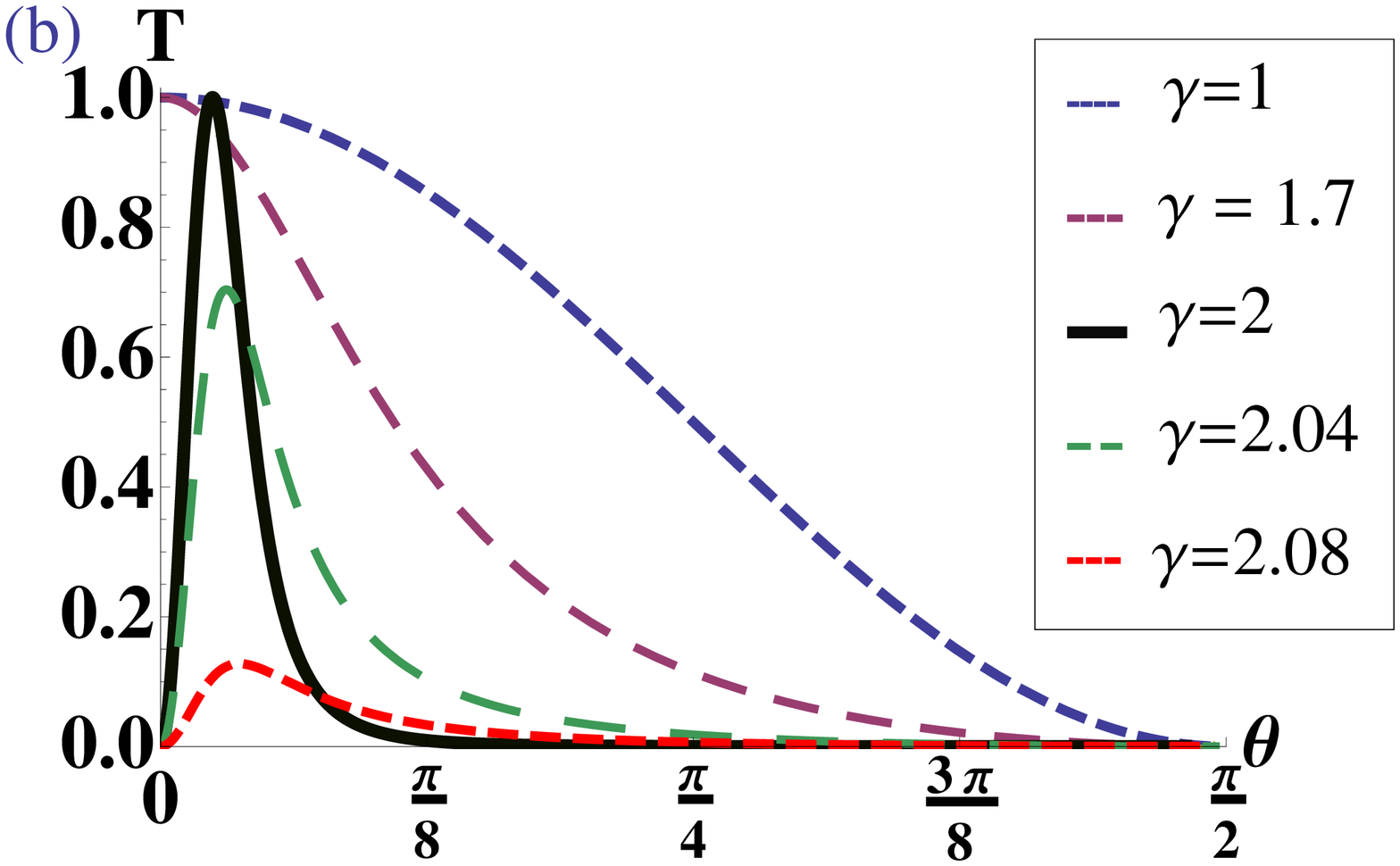}
    }%
    \caption {(color online) The transmission probability as a function of the angle of incidence, for scattering
    off a step in $y$ (a), and a step in  $x$ (b). The various lines correspond to deformations above and
    below the critical value, $\gm = 2$.}
    \label{analytic}
\end{figure}%

At normal incidence, we find a striking result: the transmission probability is exactly $1$ for all
deformations, ranging from a non-deformed lattice and up to the critical deformation. Since the waves are transferred to a region with lower index of refraction, and a single boundary exist, the result is {\it non-resonant unit transmission} in an optical system in analogy with Klein tunneling. Note that our result is not sensitive to the wavelength or the height of the step, and since there is a single boundary, the unit transmission does not result from interference. As such, the unit transmission obtained here is completely non-resonant.
This is an exceptional case in optics where other cases of unit transmission are resonance effects characterized by fine tuning of wavelength and/or potential parameters.

Another example of non-resonant unit transmission was obtained in graphene (non-deformed honeycomb) in Ref.~\cite{Ando} where the unit transmission was linked to the $\pi$ geometrical (Berry) phase accumulated by circumventing the Dirac point. Our findings do not support this linkage, since at the critical deformation the Dirac points merge and the geometrical phase vanishes. Nevertheless, the transmission probability at normal incidence is $1$, implying that it is not directly related to the geometrical phase. Moreover, we find that, for increasing $\gamma$ above $1$ the dependence of $T$ on the angle of incidence, $\theta$, decreases significantly, and at the critical deformation, $T$ is angle independent up to very large angles (Fig.\ref{analytic}a). Even though the transverse wave-vector is conserved, the transverse current changes sign, implying that the wave experiences negative refraction. Due to  extremely weak angular dependence, the reflected wave is negligible, in contrast to other systems where negative refraction is accompanied by significant reflection \cite{Ertugrul}. At deformations above the critical one, the $T(\th)$ is always smaller than $1$, but the angular dependence is still extremely weak (Fig.\ref{analytic}a).

\subparagraph{Step in $x$:} The optical potential is $V_N (x,y) = V_0 \. \Theta(x)$. Well below the critical
deformation, the leading term in the \Hm~is linear in $p_x$, and the behavior is identical to the
previous case, i.e., at normal incidence non-resonant unity transmission is obtained. However, as the
Dirac points get closer, the Dirac cones are distorted and the valleys are no longer separable.
This situation occurs when the linear term in $p_x$ is comparable to the quadratic term. As the deformation
approaches the critical one, the quadratic term becomes the leading term and the nature of the scattering
process changes dramatically: since the \Hm~is quadratic in $p_x$, there are four possible solutions for a
specified propagation constant $\beta$. Two solutions have a real wave-vector corresponding to
propagating waves, and the other two have an imaginary wave-vector, corresponding to exponentially decaying waves. The presence of the exponential waves changes the transmission probability, and at normal incidence the transmission vanishes completely, for all deformations above the critical one (Fig.(\ref{analytic}b)). This is reflected in a rapid change in $T(\th = 0)$ from unit transmission to total reflection over a very small range
in the parameters space, e.g., at $p_x a \approx 0.6$, the linear and quadratic terms are equal at $\gm \approx 1.97$, and for higher values of $\gm$ the quadratic term dominates. Therefore, we obtain unit transmission for $\gm \lesssim 1.95$, and total reflection for $\gm \gtrsim 2$.
The angular dependence is characterized by transmission peaks at small angles, so that such potential step
filters out the $p_x = 0$ mode.
\begin{figure}[t]
    \center
    {\includegraphics[width=0.80\textwidth]{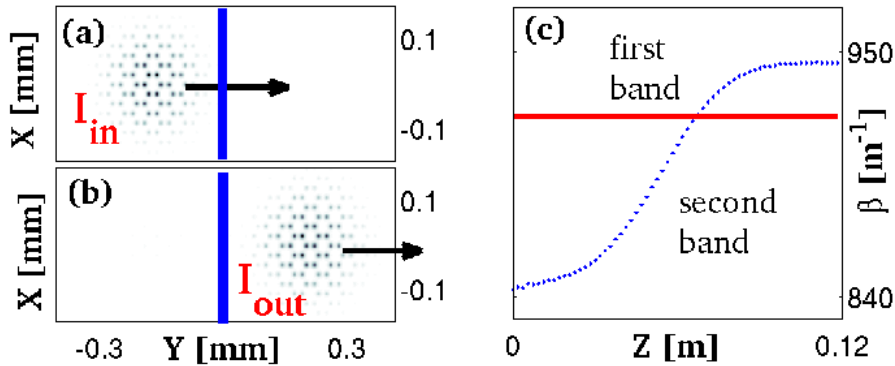}}%
    \caption {The input (a) and output (b) intensities for a step in the $y-$direction, for deformation
    close to the critical one.
    The position of the step is indicated by the vertical solid line.
    \\ (c) the mean propagation constant along the propagation (dots). The solid line indicates the
     intersection of the bands.}
    \label{y_dir}
\end{figure}%

Lattices with a deformation close to the critical one, have an effective \Hm~that resembles the \Hm~of bilayer graphene, and the scattering properties in the $x-$direction resembles that of bilayer graphene as well \cite{Katsnelson_b}. Thus, the critically deformed honeycomb is a hybrid of monolayer and bilayer graphene.

In order to supplement the analytic treatment which includes various assumptions (e.g.,
tight binding, sharp potential step), we re-examine the
scattering problem numerically, using a continuous paraxial wave equation with honeycomb photonic lattice \cite{peleg:103901,Bahat-Treidel:08} (not relying on tight binding at all).
We solve the eigenvalue problem for deformed honeycomb refractive index (Fig.\ref{lattice}b)
and find the Bloch waves of the system, where the deformation is close to the critical one.
We construct the initial wave packet from Bloch waves of the second band and propagate it using
Eq.(\ref{nls}) with the additional smooth step-like optical potential.
For an index step in the $y-$direction, we find that the entire wave packet is transmitted
to the region of lower index, manifesting non-resonant unity transmission in a 2D system (Fig.\ref{y_dir}). We calculate the mean propagation constant, $\la \hat H \ra$, during the propagation, and verify that the wave packet transforms from the second band to the first (Fig.\ref{y_dir}c). Moreover, we repeat the simulation with potentials of finite width and various shapes, and indeed, the unit transmission is independent of the shape and width of the potential.

\begin{figure}[t]
    {\includegraphics[width=0.80\textwidth]{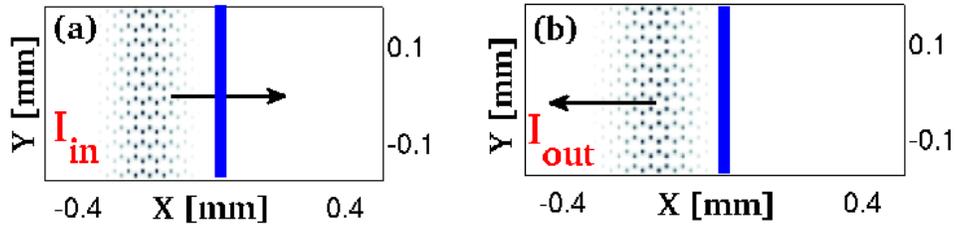}}%
    \caption {The input (a) and output (b) intensities for a step in the deformation direction.
    The position of the step is indicated by the vertical solid line.}
    \label{x_dir}
\end{figure}

As for the $x-$direction, in order to demonstrate the total reflection, one must use a very broad beam in the $y-$direction, since even waves with very small momentum in $y-$ experience significant transmission.
We preform such simulations and find that a wave-packet that is completely extended in $y$ is totally reflected
(Fig.\ref{x_dir}).

In conclusion, we have shown that the scattering of a wave packet in deformed honeycomb lattice is extremely unique:
below the critical deformation, waves that are normally incident upon a potential step in any direction exhibit non-resonant unit transmission, and the angular dependence of the transmission becomes very flat as the deformation approach the critical one. Moreover, we have demonstrated perfect Klein tunneling with zero geometrical phase rather that $\pi$. Beyond the critical deformation, the system exhibits non-resonant total reflection for waves incident normally upon a step in the deformation direction, which indicates that close to the critical deformation the system has some of the unique characteristics of both monolayer and bilayer graphene.
Unlike other systems (graphene or cold atoms in optical lattices) photonic lattices in which one can observe the field itself offer us a great opportunity to finally have a direct observation of Klein tunneling.


\begin{thebibliography}{10}

\bibitem{klein}
O. Klein, Z. Physics {\bf 53},  157  (1929).

\bibitem{Ando}
T. {Ando}, T. {Nakanishi}, and R. {Saito}, Journal of the Phys. Society of
  Japan {\bf 67},  2857  (1998).

\bibitem{Katsnelson_b}
M.~I. Katsnelson {\it et~al.}, Nature  (2006).

\bibitem{Huard-2007-98}
B. Huard {\it et~al.}, Phys. Rev. Lett. {\bf 98},  236803  (2007).

\bibitem{stander:026807}
N. Stander {\it et~al.}, Phys. Rev. Lett. {\bf 102},
  026807  (2009).

\bibitem{kim}
A.~F. Young and P. Kim, Nature Physics {\bf 5},  222  (2009).

\bibitem{sepkhanov:063813}
R.~A. Sepkhanov {\it et~al.}, Phys. Rev. A {\bf 75},
  063813  (2007).

\bibitem{sepkhanov:045122}
R.~A. Sepkhanov {\it et~al.}, Phys. Rev. B {\bf 78},
  045122  (2008).

\bibitem{peleg:103901}
O. Peleg {\it et~al.}, Phys. Rev. Lett. {\bf 98},  103901  (2007).

\bibitem{ablowitz:053830}
M.~J. Ablowitz {\it et~al.}, Phys. Rev. A {\bf 79},  053830  (2009).

\bibitem{zhu-2007-98}
S.-L. Zhu {\it et~al.}, Phys. Rev. Lett. {\bf 98},  260402  (2007).

\bibitem{Bahat-Treidel:08}
O. Bahat-Treidel {\it et~al.}, Opt. Lett. {\bf 33},  2251  (2008).

\bibitem{pereira:045401}
V.~M. Pereira {\it et~al.}, Phys. Rev. B {\bf 80},
  045401  (2009).

\bibitem{Wunsch}
B. Wunsch, F. Guinea, and F. Sols, New J. Phys  (2008).

\bibitem{dietl:236405}
P. Dietl {\it et~al.}, Phys. Rev. Lett. {\bf 100},  236405
  (2008).

\bibitem{gusynin-2007-21}
V.~P. Gusynin, S.~G. Sharapov, and J.~P. Carbotte, Int. J.
  Mod. Phys. B {\bf 21},  4611  (2007).

\bibitem{Ertugrul}
C. Ertugrul {\it et~al.}, Nature {\bf 423},  604  (2003).

\end{thebibliography}
\end{document}